\documentclass[aos,MSNbibl,seceqn,citesort,dvips]{arximspdf}
\usepackage{dcolumn}

\doi{10.1214/11-AOS919}
\volume{39}
\issue{5}
\pubyear{2011}
\firstpage{2607}
\lastpage{2625}

\makeatletter

\newcolumntype{d}[1]{D{.}{.}{#1}}

\newtheorem{theorem}{Theorem}[section]
\newtheorem{lemma}{Lemma}
\newtheorem{proposition}[theorem]{Proposition}
\newproclaim{Algorithm}{Algorithm}

\newproclaim{Remark}{Remark}

\newcommand{\mb}{\bolds}

\newcommand{\Prefix}[2]{{}#1#2}

\makeatother

\begin{document}
\begin{frontmatter}

\title{Penalized maximum likelihood estimation and variable selection
in geostatistics}
\runtitle{PMLE and variable selection in geostatistics}

\begin{aug}
\author[A]{\fnms{Tingjin} \snm{Chu}\ead[label=e1]{tingjin.chu@colostate.edu}},
\author[A]{\fnms{Jun} \snm{Zhu}\thanksref{t1}\ead[label=e2]{jzhu@stat.wisc.edu}}
\and
\author[B]{\fnms{Haonan} \snm{Wang}\corref{}\thanksref{t2}\ead[label=e3]{wanghn@stat.colostate.edu}}
\runauthor{T. Chu, J. Zhu and H. Wang}
\affiliation{Colorado State University, University
of Wisconsin, Madison,\break and Colorado State University}
\address[A]{T. Chu\\
H. Wang\\
Department of Statistics\\
Colorado State University\\
Fort Collins, Colorado 80523\\
USA\\
\printead{e1}\\
\phantom{E-mail: }\printead*{e3}}
\address[B]{J. Zhu\\
Department of Statistics\\
\quad and Department of Entomology\\
University of Wisconsin\\
Madison, Wisconsin 53706\\
USA\\
\printead{e2}} 
\end{aug}

\thankstext{t1}{Supported in part by a USDA CSREES Hatch project.}
\thankstext{t2}{Supported in part by NSF Grants DMS-07-06761,
DMS-08-54903 and DMS-11-06975, and by the Air Force Office of Scientific
Research under contract number FA9550-10-1-0241.}

\received{\smonth{7} \syear{2010}}
\revised{\smonth{8} \syear{2011}}

%
\begin{abstract}
We consider the problem of selecting covariates in spatial linear
models with Gaussian process errors. Penalized maximum likelihood
estimation (PMLE) that enables simultaneous variable selection and
parameter estimation is developed and, for ease of computation, PMLE is
approximated by one-step sparse estimation (OSE). To further improve
computational efficiency, particularly with large sample sizes, we
propose penalized maximum covariance-tapered likelihood estimation
(PMLE$_{\mathrm{T}}$) and its one-step sparse estimation (OSE$_{\mathrm T}$).
General forms of penalty functions with an emphasis on smoothly clipped
absolute deviation are used for penalized maximum likelihood.
Theoretical properties of PMLE and OSE, as well as their approximations
PMLE$_{\mathrm T}$ and OSE$_{\mathrm T}$ using covariance tapering, are
derived, including consistency, sparsity, asymptotic normality and the
oracle properties. For covariance tapering, a by-product of our
theoretical results is consistency and asymptotic normality of maximum
covariance-tapered likelihood estimates. Finite-sample properties of
the proposed methods are demonstrated in a simulation study and, for
illustration, the methods are applied to analyze two real data sets.
\end{abstract}

%

\end{frontmatter}

\section{Introduction}\label{secintro}

Geostatistical models are popular tools for the analysis of spatial
data in many disciplines. It is often of interest to estimate model
parameters based on data at sampled locations and perform spatial
interpolation (also known as Kriging) of a response variable at
unsampled locations within a spatial domain of interest
\cite{Cressie1993,Stein1999,SchabenbergerGotway2005}. In addition, a
practical issue that often arises is how to select the best model\vadjust{\goodbreak} or a
best subset of models among many competing
ones~\cite{HoetingDavisMertonThompson2006}. Here we focus on selecting
covariates in a~spatial linear model, which we believe is a problem
that is underdeveloped in both theory and methodology despite its
importance in geostatistics. The spatial linear model for a response
variable under consideration has two additive components: a fixed
linear regression term and a stochastic error term. We assume that the
error term follows a Gaussian process with mean zero and a covariance
function that accounts for spatial dependence. Our chief objective is
to develop a set of new methods for the selection of covariates and
establish their asymptotic properties. Moreover, we devise efficient
algorithms for computation, making these methods feasible for practical
usage.


For linear regression with independent errors, variable selection has
been widely studied in the literature. The more traditional methods
often involve hypothesis testing such as $F$-tests in a stepwise
selection procedure \cite{DraperSmith1998}. An alternative approach
is to select models using information discrepancy such as a
Kolmogorov--Smirnov, Kullback--Leibler or Hellinger discrepancy
\cite{LinhartZucchini1986}. In recent years, penalized methods are
becoming increasingly popular for variable selection. For example,
Tibshirani \cite{Tibshirani1996} developed a least absolute shrinkage and
selection operator (LASSO), whereas Fan and Li \cite{FanLi2001}
proposed a
nonconcave penalized likelihood method with a smoothly clipped absolute
deviation (SCAD) penalty.
Efron et al. \cite{EfronHastieJohnstoneTibshirani2004} devised least angle
regression (LARS) algorithms, which allow computing all LASSO estimates
along a~path of its tuning parameters at a low computational order.
More recently, Zou \cite{Zou2006} improved LASSO and the resulting
adaptive LASSO enjoys the oracle properties as SCAD, in terms of
selecting the true model. Zou and Li \cite{ZouLi2008} proposed one-step sparse
estimation in the nonconcave penalized likelihood approach, which
retains the oracle properties and utilizes LARS algorithms.

For spatial linear models in geostatistics, in contrast, statistical
methods for a principled selection of covariates are limited.
Hoeting et al. \cite{HoetingDavisMertonThompson2006} suggested Akaike's
information criterion (AIC) with a finite-sample correction for
variable selection. Like information-based selection in general,
computation can be costly especially when the number of covariates
and/or the sample sizes are large. Thus, these authors considered only
a subset of the covariates that may be related to the abundance of the
orange-throated whiptail lizard in southern California, in order to
make it tractable to evaluate their AIC-based model selection.
Huang and Chen \cite{HuangChen2007} developed a~model selection
criterion in
geostatistics, but for the purpose of Kriging rather than selection of
covariates. Further, Wang and Zhu \cite{WangZhu2009} proposed penalized least
squares (PLS) for a spatial linear model where the error process is
assumed to be strong mixing without the assumption of a Gaussian
process. This method includes spatial autocorrelation only indirectly
in the sense that the objective function involves a sum of squared
errors ignoring spatial dependence. A~spatial block bootstrap is then
used to account for spatial dependence when estimating the variance of
PLS estimates.\vadjust{\goodbreak}

Here we take an alternative, parametric approach and assume that the
errors in the spatial linear model follow a Gaussian process. Our main
innovation here is to incorporate spatial dependence directly into a
penalized likelihood function and achieve greater efficiency in the
resulting penalized maximum likelihood estimates (PMLE). Unlike
computation of PLS estimates which is on the same order as ordinary
least squares estimates, however, penalized likelihood function for a
spatial linear model will involve operations of a covariance matrix of
the same size as the number of observations. Thus, the computational
cost can be prohibitively high as the sample size becomes large. It is
essential that our new methods address this issue. To that end, we
utilize one-step sparse estimation (OSE) and LARS algorithms in the
computation of PMLE to gain computational efficiency. In addition, we
explore covariance tapering, which further reduces computational cost
by replacing the exact covariance matrix with a sparse one
\cite
{FurrerGentonNychaka2006,KaufmanSchervishNychka2008,DuZhangMandrekar2009}.
We establish the asymptotic properties of
both PMLE and OSE, as well as their covariance-tapered counterparts.
As a by-product, we establish new results for covariance-tapered MLE
which, to the best of our knowledge, have not been established before
and can be of independent interest.

The remainder of the paper is organized as follows. In
Section \ref{sectaper} we develop penalized maximum covariance-tapered
likelihood estimation (PMLE$_{\mathrm T}$) that enables simultaneous
variable selection and parameter estimation, as well as an
approximation of the PMLE$_{\mathrm T}$ by one-step sparse estimation
(OSE$_{\mathrm T}$) to enhance computational efficiency. PMLE and OSE are
regarded as a special case of PMLE$_{\mathrm T}$ and OSE$_{\mathrm T}$. We
establish asymptotic properties of PMLE and OSE in
Section \ref{secPMLEasymp} and those of PMLE$_{\mathrm T}$ and OSE$_{\mathrm T}$ under covariance tapering in Section \ref{sectaperasymp}. In
Section \ref{secnumeric} finite-sample properties of the proposed
methods are investigated in a simulation study and, for illustration,
the methods are applied to analyze two real data sets. We outline the
technical proofs in Appendices \ref{appA1} and \ref{appA2}.


\section{Maximum likelihood estimation: Penalization and covariance
tapering}\label{sectaper}

%
%

\subsection{Spatial linear model and maximum likelihood estimation}

For a spatial domain of interest $R$ in $\mathbb{R}^d$, we consider a
spatial process $\{y(\mathbf{s})\dvtx\mathbf{s}\in R\}$ such that
%
\begin{equation}\label{Eqnmodel}
y(\mathbf{s}) = \mathbf{x}(\mathbf{s})^{T}\mb{\beta}+\varepsilon(\mathbf{s}),
\end{equation}
where $\mathbf{x}(\mathbf{s})=(x_{1}(\mathbf{s}),\ldots,x_{p}(\mathbf{s}))^{T}$ is a
$p\times1$ vector of covariates at location $\mathbf{s}$ and $\mb{\beta
}=(\beta_{1},\ldots,\beta_{p})^{T}$ is a $p\times1$ vector of
regression coefficients. We assume that the error process $\{\varepsilon
(\mathbf{s})\dvtx\mathbf{s}\in R\}$ is a Gaussian process with mean zero and a
covariance function
%
\begin{equation}\label{Eqncovariance}
\gamma(\mathbf{s}, \mathbf{s}';\mb{\theta}) = \operatorname{cov}\{\varepsilon(\mathbf{s}),
\varepsilon(\mathbf{s}')\},
\end{equation}
where $\mathbf{s}, \mathbf{s}'\in R$ and $\mb{\theta}$ is a $q\times1$
vector\vadjust{\goodbreak}
of covariance function parameters.

Let $\mathbf{s}_1,\ldots,\mathbf{s}_{N}$ denote $N$ sampling sites in $R$. Let
$\mathbf{y}=(y(\mathbf{s}_1),\ldots,y(\mathbf{s}_{N}))^{T}$ denote an $N\times1$
vector of response variables and
$\mathbf{x}_j=(x_j(\mathbf{s}_1),\ldots,x_j(\mathbf{s}_{N}))^{T}$ denote an
$N\times1$ vector of the $j$th covariate with $j=1,\ldots,p$, at the
$N$ sampling sites. Further, let $\mathbf{X}=[\mathbf{x}_1,\ldots,\mathbf{x}_p]$
denote an $N\times p$ design matrix of covariates and
$\mb{\Gamma}=[\gamma(\mathbf{s}_i,\mathbf{s}_{i'};\mb{\theta})]_{i,i'=1}^N$
denote an $N\times N$ covariance matrix. In this paper, we consider
general forms for the the covariance matrix $\mb{\Gamma}$ and describe
suitable regularity conditions in Sections \ref{secPMLEasymp} and
\ref{sectaperasymp}. By (\ref{Eqnmodel}) and~(\ref{Eqncovariance}),
we have
%
\begin{equation}\label{EqnGP}
\mathbf{y}\sim N(\mathbf{X}\mb{\beta},\mb{\Gamma}).
\end{equation}

Let $\mb{\eta}=(\mb{\beta}^T,\mb{\theta}^T)^T$ denote a $(p+q)\times1$
vector of model parameters consisting of both regression coefficients
$\mb{\beta}$ and covariance function parameters~$\mb{\theta}$. By
(\ref{EqnGP}), the log-likelihood function of $\mb{\eta}$ is
%
\begin{eqnarray}\label{Eqnloglike}
\ell(\mb{\eta};\mathbf{y},\mathbf{X}) &=& -(N/2)\log(2\pi)-{(1/2)\log}|\mb{\Gamma
}|\nonumber\\[-8pt]\\[-8pt]
&&{}-(1/2)(\mathbf{y}-\mathbf{X}\mb{\beta})^T\mb{\Gamma}^{-1}(\mathbf{y}-\mathbf{X}\mb
{\beta}).\nonumber
\end{eqnarray}
Let $\widehat{\mb{\eta}}_{\mathrm{MLE}}=\arg\max_{\mb{\eta}}\{\ell
(\mb{\eta};\mathbf{y},\mathbf{X})\}$ denote the maximum likelihood estimate
(MLE) of $\mb{\eta}$.

\subsection{Covariance tapering and penalized maximum likelihood}

It is well known that computation of MLE for a spatial linear model is
of order $N^3$ and can be very demanding when the sample size $N$
increases \cite{Cressie1993}. There are various approaches to
alleviating the computational cost. Here we consider covariance
tapering, which could effectively reduce our computational cost in
practice. Furrer et al. \cite{FurrerGentonNychaka2006} considered
tapering for
Kriging and demonstrated that not only tapering enhances computational
efficiency but also achieves asymptotically optimality in terms of mean
squared prediction errors under infill asymptotics. For parameter
estimation via maximum likelihood,
Kaufman et al. \cite{KaufmanSchervishNychka2008} established
consistency of
tapered MLE, whereas Du et al. \cite{DuZhangMandrekar2009} established the
asymptotic distribution, also under infill asymptotics. However, both
Kaufman et al. \cite{KaufmanSchervishNychka2008} and
Du et al. \cite{DuZhangMandrekar2009} focused on the parameters in the
Mat\'{e}rn family of covariance functions and did not consider
estimation of the regression coefficients. In contrast, our primary
interest is in the estimation of regression coefficients and we
investigate the asymptotic properties under increasing domain
asymptotics, which, to the best of our knowledge, have not been
established in the literature before.

Recall that $\mb{\Gamma}=[\gamma(\mathbf{s}_i,\mathbf{s}_{i'})]_{i,i'=1}^{N}$
is the covariance matrix of $\mathbf{y}$. Assuming second-order
stationarity and isotropy, we let
$
\gamma(d)=\gamma(\mathbf{s},\mathbf{s}'),
$
where $d=\|\mathbf{s}-\mathbf{s}'\|$ is the lag distance between two sampling
sites $\mathbf{s}$ and $\mathbf{s}'$ in $R$. Let $K_{\mathrm T}(d,\omega)$ denote a
tapering function, which is an isotropic autocorrelation function when
$0<d<\omega$ and $0$ when $d \geq\omega$, for a given threshold
distance $\omega>0$. Compactly supported correlation functions can be
used as the tapering functions \cite{Wendland1995}. For example,
%
\begin{equation}\label{eqnKT}
K_{\mathrm T}(d, \omega)=(1-d/\omega)_+,
\end{equation}
where $x_+=\mbox{max}\{x,0\}$, in which case the correlation is 0 at
lag distance greater than the threshold\vspace*{2pt} distance $\omega$. Let $\mb
{\Delta}(\omega)=[K_{\mathrm T}(d_{ii'},\omega)]_{i,i'=1}^{N}$ denote an
${N}\times{N}$ tapering matrix. Then a tapered covariance matrix of
$\mb{\Gamma}$ is defined as
$
\mb{\Gamma}_{\mathrm T}=\mb{\Gamma}\circ\mb{\Delta}(\omega),
$
where $\circ$ denotes the Hadamard product (i.e., elementwise product).

We approximate the log-likelihood function by replacing $\mb{\Gamma}$
in (\ref{Eqnloglike}) with the tapered covariance matrix $\mb{\Gamma
}_{\mathrm T}$ and obtain a covariance-tapered log-likelihood function
%
\begin{eqnarray}\label{EqnTaplog}
\ell_{\mathrm
T}(\mb{\eta};\mathbf{y},\mathbf{X})&=&-({N}/2)\log(2\pi)-{(1/2)\log}|\mb
{\Gamma}_{\mathrm T}|\nonumber\\[-8pt]\\[-8pt]
&&{}-(1/2)(\mathbf{y}-\mathbf{X}\mb{\beta})^T\mb{\Gamma}_{\mathrm
T}^{-1}(\mathbf{y}-\mb
{X}\mb{\beta}).\nonumber
\end{eqnarray}
We let $\widehat{\mb{\eta}}_{\mathrm{MLE}_{\mathrm{T}}}=\arg\max_{\mb{\eta}}\{\ell
_{\mathrm T}(\mb{\eta};\mathbf{y},\mathbf{X})\}$ denote the maximum
covariance-tapered likelihood estimate (MLE$_{\mathrm T}$) of $\mb{\eta}$.

Let $\mb{\Gamma}_{k,\mathrm{T}}=\partial\mb{\Gamma}_{\mathrm
T}/\partial\theta_k=\mb{\Gamma}_{k} \circ\mb{\Delta}(\omega)$,
$\mb{\Gamma}_{\mathrm T}^k=\partial\mb{\Gamma}_{\mathrm
T}^{-1}/\partial \theta_k=\mb{\Gamma}^{k} \circ\mb{\Delta}(\omega)$,
$\mb{\Gamma}_{kk',\mathrm T}=\partial^2 \mb{\Gamma}_{\mathrm
T}/\partial \theta_k \,\partial\theta_{k'}=\mb{\Gamma}_{kk'} \circ
\mb{\Delta}(\omega)$, and $\mb{\Gamma}_{\mathrm T}^{kk'}=\partial^2
\mb{\Gamma}_{\mathrm T}^{-1}/\partial\theta_k \,\partial
\theta_{k'}=\mb{\Gamma}^{kk'} \circ\mb{\Delta}(\omega)$ denote the
covariance-tapered version of $\mb{\Gamma}_{k}$, $\mb{\Gamma}^k$,
$\mb{\Gamma}_{kk'}$ and $\mb{\Gamma}^{kk'}$, respectively. From
(\ref{EqnTaplog}), $\ell'_{\mathrm T}(\mb{\beta}) =
\mathbf{X}^T\mb{\Gamma}_{\mathrm
T}^{-1}(\mathbf{y}-\mathbf{X}\mb{\beta})$ and the $k$th element of
$\ell'_{\mathrm T}(\mb{\theta})$ is\break
$-(1/2)\operatorname{tr}(\mb{\Gamma}_{\mathrm T}^{-1}\mb{\Gamma}_{k,\mathrm
T})-(1/2)(\mathbf{y}-\mathbf{X}\mb{\beta})^T\mb{\Gamma}_{\mathrm
T}^k(\mathbf{y}-\mathbf{X}\mb{\beta})$. Moreover, $\ell''_{\mathrm
T}(\mb{\beta},\mb{\beta}) = -\mathbf{X}^T\mb{\Gamma}_{\mathrm
T}^{-1}\mathbf{X}$, the $k$th column of $\ell''_{\mathrm
T}(\mb{\beta},\mb{\theta})$ is $\mathbf{X}^T\mb{\Gamma}_{\mathrm
T}^k(\mathbf{y}-\mathbf{X}\mb{\beta})$, and the $(k,k')$th entry of
$\ell''_{\mathrm T}(\mb{\theta},\mb{\theta})$ is
$-(1/2)\{\operatorname{tr}(\mb{\Gamma}_{\mathrm T}^{-1}\mb{\Gamma}_{kk',\mathrm
T}+\mb{\Gamma}_{\mathrm T}^k\mb{\Gamma}_{k',\mathrm
T})+(\mathbf{y}-\mathbf{X}\mb{\beta})^T\mb{\Gamma}_{\mathrm T}^{kk'}
(\mathbf{y}-\mathbf{X}\mb{\beta})\}$. Since $E\{-\ell''_{\mathrm
T}(\mb{\beta}$, $\mb{\theta})\}= \mb{0}$, the covariance-tapered
information matrix of $\mb{\eta}$ is $\mathbf{I}_{\mathrm
T}(\mb{\eta})=\operatorname{diag}\{\mathbf{I}_{\mathrm
T}(\mb{\beta})$, $\mathbf{I}_{\mathrm T}(\mb{\theta})\}$, where
$\mathbf{I}_{\mathrm T}(\mb{\beta}) = E\{-\ell''_{\mathrm
T}(\mb{\beta},\mb{\beta})\} = \mathbf{X}^T\mb{\Gamma}_{\mathrm
T}^{-1}\mathbf{X}$ and the $(k,k')$th entry of $\mathbf{I}_{\mathrm
T}(\mb{\theta})=E\{-\ell''_{\mathrm T}(\mb{\theta},\mb{\theta})\}$ is
$t_{kk',\mathrm T}/2$ with $t_{kk',\mathrm T} =
\operatorname{tr}(\mb{\Gamma}_{\mathrm T}^{-1}\mb{\Gamma}_{k,\mathrm
T}\mb{\Gamma}_{\mathrm T}^{-1}\mb{\Gamma}_{k',\mathrm T}) =
\operatorname{tr}(\mb{\Gamma}_{\mathrm T}\mb{\Gamma}_{\mathrm
T}^k\times\allowbreak\mb{\Gamma}_{\mathrm T}\mb{\Gamma}_{\mathrm T}^{k'})$.

Now, we define a covariance-tapered penalized log-likelihood function as
%
\begin{equation}\label{EqnTapQ}
Q_{\mathrm T}(\mb{\eta})=\ell_{\mathrm T}(\mb{\eta};\mathbf{y},\mathbf{X})-N\sum
_{j=1}^{p}p_{\lambda}(|\beta_{j}|),
\end{equation}
where $\ell_{\mathrm T}(\mb{\eta};\mathbf{y},\mathbf{X})$ is a
covariance-tapered log-likelihood function as defined
in~(\ref{EqnTaplog}). Moreover, we let $\widehat{\mb{\eta}}_{\mathrm
{PMLE}_{\mathrm T}}=\arg\max_{\mb{\eta}}\{Q_{\mathrm T}(\mb{\eta})\}$
denote the penalized maximum covariance-tapered likelihood estimate
(PMLE$_{\mathrm T}$) of $\mb{\eta}$.



For penalty functions, we mainly consider smoothly clipped absolute
deviation (SCAD) defined as
%
\begin{equation}\label{EqnSCAD}
p_\lambda(\beta) = \cases{
\lambda|\beta|, \qquad\hspace*{34.5pt} \mbox{if $|\beta| \leq\lambda$},\vspace*{2pt}\cr
\lambda^2+(a-1)^{-1}(a\lambda|\beta|-\beta^2/2-a\lambda^2+\lambda^2/2),\vspace*{2pt}\cr
\hspace*{81pt}\mbox{if $\lambda<|\beta| \leq a\lambda$},\vspace*{2pt}\cr
(a+1)\lambda^2/2, \qquad \mbox{if $|\beta| > a\lambda$},}
\end{equation}
for some $a>2$ \cite{Fan1997}. For i.i.d. error in standard linear
regression, variable selection and parameter estimation under the SCAD
penalty are shown to possess three desirable properties: unbiasedness,
sparsity and continuity \cite{FanLi2001}. For spatial linear
regression (\ref{Eqnmodel}), these properties continue to hold for the
SCAD penalty following arguments similar to those in
Wang and Zhu~\cite{WangZhu2009}.\looseness=1

To compute PMLE$_{\mathrm T}$ under the SCAD penalty, Fan and Li \cite
{FanLi2001} proposed a~locally quadratic approximation (LQA) of the
penalty function and a~New\-ton--Raphson algorithm. Although fast, a
drawback of the LQA algorithm is that once a regression coefficient is
shrunk to zero, it remains to be zero in the remainder iterations. More
recently, Zou and Li \cite{ZouLi2008} developed a unified algorithm to
improve computational efficiency, which, unlike the LQA algorithm, is
based on the locally linear approximation (LLA) of the penalty
function. Moreover, Zou and Li \cite{ZouLi2008} proposed one-step LLA
estimation that approximates the solution after just one iteration in a
Newton--Raphson-type algorithm starting at the MLE. We extend this
one-step LLA estimation to approximate PMLE$_{\mathrm T}$ for the spatial
linear model as follows.

\begin{Algorithm}\label{Algorithm1}
At the initialization step, we let $\mb{\eta}_{\mathrm T}^{(0)}=\widehat{\mb{\eta}}_{\mathrm{MLE_T}}$ with $\mb{\beta}_{\mathrm T}^{(0)}=\widehat{\mb{\beta}}_{\mathrm{MLE_T}}$ and
$\mb{\theta}^{(0)}_{\mathrm T}=\widehat{\mb{\theta}}_{\mathrm{MLE}_{\mathrm{T}}}$. We
then update $\mb{\beta}$ by maximizing
%
\begin{equation}\label{EqnTapLLA}\qquad
Q^*_{\mathrm T}(\mb{\beta})=-(1/2)(\mathbf{y}-\mathbf{X}\mb{\beta})^T\mb{\Gamma
}_{\mathrm T}\bigl(\mb{\theta}_{\mathrm T}^{(0)}\bigr)^{-1}(\mathbf{y}-\mathbf{X}\mb{\beta})-N\sum
_{j=1}^{p}p'_{\lambda}\bigl(\bigl|\beta^{(0)}_{j{\mathrm T}}\bigr|\bigr)|\beta_j|
\end{equation}
with respect to $\mb{\beta}$, where the first term is from (\ref
{EqnTaplog}) and the second term is an LLA of the penalty function in
(\ref{EqnTapQ}). The resulting one-step sparse estimate (OSE) of $\mb
{\beta}$ is denoted as $\widehat{\mb{\beta}}_{\mathrm{OSE}_{\mathrm T}}$. We may also
update $\mb{\theta}$ by maximizing~(\ref{EqnTaplog}) with respect to
$\mb{\theta}$ given $\widehat{\mb{\beta}}_{\mathrm{OSE}_{\mathrm T}}$. The resulting
OSE of $\mb{\theta}$ is denoted as $\widehat{\mb{\theta}}_{\mathrm{OSE}_{\mathrm T}}$.
We let $\widehat{\mb{\eta}}_{\mathrm{OSE}_{\mathrm T}}=(\widehat{\mb{\beta}}_{\mathrm{
OSE}_{\mathrm T}}^T, \widehat{\mb{\theta}}_{\mathrm{OSE}_{\mathrm T}}^T)^T$ denote the
OSE$_{\mathrm T}$ of $\mb{\eta}$, which approximates~%
$\widehat{\mb{\eta}}_{\mathrm{PMLE}_{\mathrm{T}}}$.\vspace*{1pt}

It is worth mentioning an alternative covariance-tapered log-likelihood
function~\cite{KaufmanSchervishNychka2008},
%
\begin{eqnarray}\label{EqnTaplog2}
  \ell_{\mathrm{T}2}(\mb{\eta};\mathbf{y},\mathbf{X})&=&-({N}/2)\log(2\pi
)-{(1/2)\log}|\mb{\Gamma}_{\mathrm T}|\nonumber\\[-8pt]\\[-8pt]
&&{} -(1/2)(\mathbf{y}-\mathbf{X}\mb{\beta})^T\{\mb{\Gamma}_{\mathrm
T}^{-1}\circ\mb
{\Delta}(\omega)\}(\mathbf{y}-\mathbf{X}\mb{\beta}).\nonumber
\end{eqnarray}
If the alternative covariance tapering is used in Algorithm
\ref{Algorithm1}, the resulting estimates of parameters, especially the
range parameter, tend to be more accurate, but require more time to
compute $\mb{\Gamma}_{\mathrm T}^{-1}\circ\mb{\Delta}(\omega)$ than
$\mb{\Gamma}_{\mathrm T}^{-1}$. For a~numerical comparison, see Section
6.1 in Chu et al. \cite{ChuZhuWang2011}.

%

Finally, two tuning parameters, $\lambda$ and $a$, in the SCAD penalty
(\ref{EqnSCAD}) need to be estimated. For computational ease, we fix
$a=3.7$ as recommended by Fan and Li~\cite{FanLi2001}. To
determine\vadjust{\goodbreak}
$\lambda$, we use the Bayesian information criterion (BIC); see
Wang et al. \cite{WangLiTsai2007b}. In particular, let
%
\begin{equation}
\widehat\sigma^2(\lambda)={N}^{-1}\{\mathbf{y}-\mathbf{X}\widehat{\mb{\beta
}}(\lambda)\}^T\mb{\Gamma}\{\widehat{\mb{\theta}}(\lambda)\}^{-1}\{\mathbf
{y}-\mathbf{X}\widehat{\mb{\beta}}(\lambda)\},
\end{equation}
where $\widehat{\mb{\beta}}(\lambda)$ and $\widehat{\mb{\theta}}(\lambda
)$ are the PMLE obtained for a given $\lambda$, and let
%
\begin{equation}
\mbox{BIC}(\lambda)={N}\log\{\widehat\sigma^2(\lambda)\}+k(\lambda)\log({N}),
\end{equation}
where $k(\lambda)$ is the number of nonzero regression coefficients
\cite{WangLiTsai2007a}.
Thus, an estimate of $\lambda$ is $\widehat{\lambda}=
\arg\min_{\lambda}\{\mbox{BIC}(\lambda)\}$.

When $\mb{\Delta}(\omega)$ is a matrix of 1's, $\mb{\Gamma}_{\mathrm T}=\mb{\Gamma}$ and $\ell_{\mathrm T}(\cdot)=\ell(\cdot)$. Similarly, we
henceforth obtain needed counterparts of the notation in this section
under maximum likelihood without covariance tapering by omitting $\mathrm
T$. For details regarding such notation, see Section 2 of
Chu et al. \cite{ChuZhuWang2011}.
\end{Algorithm}

\section{Asymptotic properties of PMLE and OSE}\label{secPMLEasymp}

%
%
\subsection{Notation and assumptions}\label{subsecnotation}

We let $\mb{\beta}_0=(\beta_{10},\ldots,\beta_{p0})^{T}=(\mb{\beta
}_{10}^{T},\mb{\beta}_{20}^{T})^{T}$ denote the true regression
coefficients, where without\vadjust{\goodbreak} loss of generality $\mb{\beta}_{10}$ is an
$s\times1$ vector of nonzero regression coefficients and $\mb{\beta
}_{20}=\mb{0}$ is a $(p-s)\times1$ zero vector. Let $\mb{\theta}_0$
denote the vector of true covariance function parameters.

We consider the asymptotic framework in Mardia and Marshall \cite
{MardiaMarshall1984} and let $n$ denote the stage of the asymptotics.
In particular, write $R_n=R$, $N_n=N$, and $\lambda_n=\lambda$.
Furthermore, define $a_{n}=\max_{1\leq j\leq p}\{|p'_{\lambda
_{n}}(|\beta_{j0}|)|\dvtx\beta_{j0}\neq0\}$ and $b_{n}=\max_{1\leq j\leq
p}\{|p''_{\lambda_{n}}(|\beta_{j0}|)|\dvtx\beta_{j0}\neq0\}$. Also, let
$\mb{\phi}(\mb{\beta})=(p^\prime_{\lambda}(|\beta_1|)\operatorname{sgn}(\beta
_1),\break \ldots,\allowbreak p'_{\lambda}(|\beta_p|)\operatorname{sgn}(\beta_p))^{T}$
and $\mb{\Phi}(\mb{\beta})=\operatorname{diag} \{p''_{\lambda}(|\beta
_1|),\ldots,p''_{\lambda}(|\beta_p|)\}$. Moreover, denote $\mb{\phi
}_n(\mb{\beta})=\mb{\phi}(\mb{\beta})$ and $\mb{\Phi}_n(\mb{\beta})=\mb
{\Phi}(\mb{\beta})$, both evaluated at $\lambda_n$. For all other
quantities that depend on $n$, the stage $n$ will be in either the left
superscript or the right subscript.

Recall that
$\Prefix^{n}{t_{kk'}}=\operatorname{tr}(\Prefix^{n}{\mb{\Gamma}}^{-1}\Prefix
^{n}{\mb{\Gamma}}_k\Prefix^{n}{\mb{\Gamma}}^{-1}\Prefix^{n}{\mb{\Gamma}}_{k'})$.
Let $\mu_{1} \leq\cdots\leq\mu_{N_n}$ denote the eigenvalues of
$\Prefix^{n}{\mb{\Gamma}}$. For $l=1,\ldots,N_n$, let $\mu^{k}_{l}$
denote the eigenvalues of $\Prefix^{n}{\mb{\Gamma}}_{k}$ such that
$|\mu^{k}_{1}| \leq\cdots\leq|\mu^{k}_{N_n}|$ and let
$\mu^{kk'}_{l}$ denote the eigenvalues of
$\Prefix^{n}{\mb{\Gamma}}_{kk'}$ such that $|\mu^{kk'}_{1}| \leq\cdots
\leq|\mu^{kk'}_{N_n}|$.

For an $N_n\times N_n$ matrix $\mathbf{A}=(a_{ij})_{i,j=1}^{N_n}$, the
Frobenius, max and spectral norm are defined\vspace*{1pt} as $\Vert\mathbf{A} \Vert
_{F}=(\sum_{i=1}^{N_n}\sum_{j=1}^{N_n}a_{ij}^2)^{1/2}$,
$\Vert\mathbf{A} \Vert_{\max}=\max\{|a_{ij}|\dvtx i,j=1,\ldots,N_n\}
$ and $\Vert\mathbf{A} \Vert_{s}=\max\{|\mu_l(\mathbf{A})
|\dvtx l=1,\ldots,N_n\}$, where $\mu_l(\mathbf{A})$ is the $l$th eigenvalue
of~$\mathbf{A}$.\vspace*{1pt}

The following regularity conditions are assumed for Theorems \ref
{thmexist} and \ref{thmose}:

\begin{longlist}[(A.11)]
\item[(A.1)]
For $\mb{\theta}\in\Omega$ where $\Omega$ is an open subset of $\mathbb
{R}^q$ such that $\mb{\eta}\in\mathbb{R}^p\times\Omega$, the covariance
function $\gamma(\cdot,\cdot;\mb{\theta})$ is twice differentiable with
respect to $\mb{\theta}$ with continuous second-order derivatives and
is positive definite in the sense that, for any $N_n\geq1$ and $
\mathbf{s}_1,\ldots,\mathbf{s}_{N_n}$, the covariance matrix $\mb{\Gamma}=[\gamma
(\mathbf{s}_i,\mathbf{s}_j;\mb{\theta})]_{i,j=1}^{N_n}$ is positive definite.

\item[(A.2)] There exist positive constants $C$, $C_k$ and $C_{kk'}$,
such that\break
$\lim_{n \rightarrow\infty} \mu_{N_n} = C<\infty$, $\lim_{n
\rightarrow\infty} |\mu_{N_n}^{k}| = C_{k}<\infty$, $\lim_{n
\rightarrow\infty} |\mu_{N_n}^{kk'}| = C_{kk'}<\infty$ for all
$k,k'=1,\ldots,q$.

\item[(A.3)]
For some $\delta>0$, there exist positive constants $D_k$, $D_{kk'}$
and $D_{kk'}^*$ such that
(i) $\Vert\Prefix^{n}{\mb{\Gamma}}_k\Vert_F^{-2}=D_k N_n^{-1/2-\delta}$
for $k=1,\ldots,q$;
(ii) either $\Vert\Prefix^{n}{\mb{\Gamma}}_k+\Prefix^{n}{\mb{\Gamma
}}_{k'}\Vert_F^{-2}=D_{kk'} N_n^{-1/2-\delta}$ or $\Vert\Prefix^{n}{\mb
{\Gamma}}_k-\Prefix^{n}{\mb{\Gamma}}_{k'}\Vert_F^{-2}=D_{kk'}^*
N_n^{-1/2-\delta}$ for any $k\neq k'$.

\item[(A.4)]
For any $k,k'=1,\ldots,q$,
(i) $\Prefix^{n}{a_{kk'}}=\lim_{n\rightarrow\infty}\{\Prefix
^{n}{t_{kk'}}(\Prefix^{n}{t_{kk}}\Prefix^{n}{t_{k'k'}})^{-1/2}\}$
exists and $\mathbf{A}_n=(\Prefix^{n}{a_{kk'}})_{k, k'=1}^q$ is nonsingular;
(ii) $|\Prefix^{n}{t_{kk}}\Prefix^{n}{t_{k'k'}}^{-1} |$
and\break $|\Prefix^{n}{t_{k'k'}}\Prefix^{n}{t_{kk}}^{-1} |$ are bounded.

\item[(A.5)]
The design matrix $\mathbf{X}$ has full rank $p$ and is uniformly bounded
in max norm with $\lim_{n\rightarrow\infty} (\mathbf{X}^{T}\mathbf{X})^{-1}=\mb{0}$.

\item[(A.6)] There exists a positive constant $C_0$, such that
$\Vert\Prefix^{n}{\mb{\Gamma}}^{-1}\Vert_{s}<C_0<\infty$.

\item[(A.7)]
For $\mb{\beta}\in\mathbb{R}^p$ and $\mb{\theta}\in\Omega$, $N_n^{-1}\mathbf
{I}_n(\mb{\beta})\rightarrow\mathbf{J}(\mb{\beta})$ and $N_n^{-1}\mathbf
{I}_n(\mb{\theta})\rightarrow\mathbf{J}(\mb{\theta})$ as $n\rightarrow
\infty$.

\item[(A.8)]
$a_{n}=O(N_n^{-1/2})$ and $b_{n}\rightarrow0$ as $n\rightarrow\infty$.

\item[(A.9)]
There exist positive constants $c_1$ and $c_2$ such that, when $\beta
_1$, $\beta_2>c_1\lambda_n$, $|p''_{\lambda_n}(\beta_1)-p''_{\lambda
_n}(\beta_2)|\leq c_2|\beta_1-\beta_2|$.

\item[(A.10)]
$\lambda_n\rightarrow0, N_n^{1/2}\lambda_n\rightarrow\infty$ as
$n\rightarrow\infty$.

\item[(A.11)]
$\liminf_{n\rightarrow\infty}\liminf_{\beta\rightarrow0^{+}}\lambda
_{n}^{-1}p'_{\lambda_{n}}(\beta)>0$.
\end{longlist}

Conditions (A.2), (A.3)(i), (A.4)(i) and (A.5) are assumed in Mardia
and Marshall \cite{MardiaMarshall1984}. Conditions (A.1) and (A.5) are
standard assumptions for MLE, whereas (A.2), (A.3)(i), (A.4)(i) and
(A.6) ensure smoothness, growth and convergence of the information
matrix \cite {MardiaMarshall1984}. Together with (A.7), they yield a
central limit theorem of $\ell'(\mb{\eta})$ and convergence in
probability of~$\ell ''(\mb{\eta})$. For establishing Theorems
\ref{thmexist} and~\ref{thmose}, only the parts (i) of (A.3) and (A.4)
are used. Moreover, the implicit asymptotic framework is increasing the
domain, where the sample size $N_n$ grows at the increase of the
spatial domain~$R_n$~\cite {MardiaMarshall1984}. Finally, (A.8)--(A.11)
are mild regularity conditions regarding the penalty function and are
sufficient for Theorems \ref{thmexist} and~\ref{thmose} to hold~%
\cite{FanLi2001} and~\cite{FanPeng2004}.

%
%
\subsection{Consistency and asymptotic normality of PMLE}


\begin{theorem}
\label{thmexist}
Under \textup{(A.1)--(A.9)}, there exists, with probability
tending to one, a local maximizer $\Prefix^{n}{\widehat{\mb{\eta}}}$ of
$Q(\mb{\eta})$
such that $\Vert \Prefix^{n}{\widehat{\mb{\eta}}}-\mb{\eta}_{0}\Vert
=O_{p}(N_n^{-1/2}+a_{n})$.\vspace*{1pt} If, in addition, \textup{(A.10)--(A.11)}
hold, then
$\Prefix^{n}{\widehat{\mb{\eta}}}=(\Prefix^{n}{\widehat{\mb{\beta
}}}{}^{T}_{1},\Prefix^{n}{\widehat{\mb{\beta}}}{}^{T}_{2},\Prefix
^{n}{\widehat{\mb{\theta}}}{}^{T})^{T}$ satisfies:
\begin{longlist}
\item Sparsity: $\Prefix^{n}{\widehat{\mb{\beta}}}_{2}=\mb{0}$
with probability tending to 1.
\item Asymptotic normality:
\begin{eqnarray*}
&&N_n^{1/2}\{\mathbf{J}(\mb{\beta}_{10})+\mb{\Phi}_n(\mb{\beta}_{10})\}
[\Prefix^{n}{\widehat{\mb{\beta}}}_{1}-\mb{\beta}_{10}+\{\mathbf{J}(\mb
{\beta}_{10})+\mb{\Phi}_n(\mb{\beta}_{10})\}^{-1}\mb{\phi}_{n}(\mb{\beta
}_{10})]
\\
&&\qquad\stackrel{D}{\longrightarrow} N(\mb{0},\mathbf{J}(\mb{\beta}_{10})), \\
&&N_n^{1/2}(\Prefix^{n}{\widehat{\mb{\theta}}}-\mb{\theta}_{0})
\stackrel{D}{\longrightarrow} N(\mb{0},\mathbf{J}(\mb{\theta}_0)^{-1}),
\end{eqnarray*}
where $\mathbf{J}(\mb{\beta}_{10})$ and $\mb{\Phi}_n(\mb{\beta}_{10})$
consist of the first $s\times s$ upper-left submatrix of $\mathbf{J}(\mb
{\beta}_0)$ and $\mb{\Phi}_n(\mb{\beta}_0)$, respectively.
\end{longlist}
\end{theorem}

Theorem \ref{thmexist} establishes the asymptotic properties of PMLE.
Under (A.1)--(A.9), there exists a local maximizer converging to the
true parameter at the rate $O_{p}(N_n^{-1/2}+a_{n})$. Since
$a_{n}=O(N_n^{-1/2})$ from (A.8), the local maximizer is root-$N_n$
consistent. As shown in Fan and Li \cite{FanLi2001}, the SCAD penalty
function satisfies (A.8)--(A.11) by choosing an appropriate tuning
parameter $\lambda_n$. Therefore, by Theorem \ref{thmexist}, the PMLE
under the SCAD penalty possesses the sparsity property and asymptotic
normality. Moreover, when the sample size $N_n$ is sufficiently large,
$\mb{\Phi}_n(\mb{\beta}_{10})$ will be close to zero. That is,
performance of the PMLE is asymptotically as efficient as the MLE of
$\mb{\beta}_1$ when knowing $\mb{\beta}_2=\mb{0}$. The arguments above
hold for other penalty functions such as $L_q$ penalty with $q<1$, but
not $q=1$.

%
%
\subsection{Consistency and asymptotic normality of OSE}

%
\begin{theorem}
\label{thmose}
Suppose that the initial value $\Prefix^{n}{\mb{\eta}^{(0)}}$ satisfies
$\Prefix^{n}{\mb{\eta}^{(0)}}-\mb{\eta}_{0}=O_p(N_n^{-1/2})$. For the
SCAD penalty,\vspace*{2pt} under \textup{(A.1)--(A.7)} and \textup{(A.10)},
the OSE $\Prefix^{n}{\widehat{\mb{\eta}}}_{\mathrm{OSE}}=(\Prefix
^{n}{\widehat{\mb{\beta}}}{}^{T}_{1,{\mathrm{OSE}}},\Prefix^{n}{\widehat{\mb
{\beta}}}{}^{T}_{2,{\mathrm{OSE}}},\Prefix^{n}{\widehat{\mb{\theta}}}{}^{T}_{
\mathrm{OSE}})^{T}$ satisfies:
\begin{longlist}
\item Sparsity: $\Prefix^{n}{\widehat{\mb{\beta}}}_{2,{
\mathrm{OSE}}}=\mb{0}$ with probability tending to 1.
\item Asymptotic normality:
\begin{eqnarray*}
N_n^{1/2}(\Prefix^{n}{\widehat{\mb{\beta}}}_{1,{\mathrm{OSE}}}-\mb{\beta
}_{10})&\stackrel{D}{\longrightarrow}& N(\mb{0},\mathbf{J}(\mb{\beta
}_{10})^{-1}),\\
N_n^{1/2}(\Prefix^{n}{\widehat{\mb{\theta}}}_{\mathrm{OSE}}-\mb{\theta
}_{0})&\stackrel{D}{\longrightarrow}& N(\mb{0},\mathbf{J}(\mb{\theta}_0)^{-1}),
\end{eqnarray*}
where $\mathbf{J}(\mb{\beta}_{10})$ consists of the first $s\times s$
upper-left submatrix of $\mathbf{J}(\mb{\beta}_0)$.
\end{longlist}
\end{theorem}

Theorem \ref{thmose} establishes the asymptotic properties of OSE such
that the OSE is sparse and asymptotically normal under the SCAD penalty.
The OSE for $\mb{\beta}_1$ and $\mb{\theta}$ has the same limiting
distribution as the PMLE and thus achieves the same efficiency. In
fact, Theorem \ref{thmose} holds for another general class of penalty
functions such that $p'_{\lambda_n}(\cdot)=\lambda_np(\cdot)$ where
$p'(\cdot)$ is continuous on $(0,\infty)$, and there is some $\alpha>0$
such that $p'(\beta)=O(\beta^{-\alpha})$ as $\beta\rightarrow{0+}$~%
\cite{ZouLi2008}. Following similar arguments for the SCAD penalty in
our Theorem \ref{thmose} and those in Zou and Li \cite{ZouLi2008}, it
can be shown that, if $N_n^{(1+\alpha)/2}\lambda_n\rightarrow\infty$
and $N_n^{1/2}\lambda_n\rightarrow0$, Theorem \ref{thmose} continues
to hold. In practice, we set the initial value $\Prefix^{n}{\mb{\eta
}^{(0)}}$ to be the MLE $\Prefix^{n}{\widehat{\mb{\eta}}}_{\mathrm{MLE}}$,
as it satisfies the consistency condition.

\section{Asymptotic properties under covariance tapering}\label{sectaperasymp}

%
%
\subsection{Notation and assumptions}

In order to establish the asymptotic properties under covariance
tapering, we continue to assume (A.1)--(A.11). We now restrict our
attention to a second-order stationary error process in~$\mathbb{R}^2$
with an isotropic covariance function $\gamma(d)$, where $d \geq0$ is
the lag distance. We also assume that the distance between any two
sampling sites is greater than a constant \cite{MardiaMarshall1984}. As
for the tapering function, we consider (\ref{eqnKT}).

Let $\gamma_k(d)=\partial\gamma(d)/\partial\theta_k$, $\gamma
_{kk'}(d)=\partial^2 \gamma(d)/\partial\theta_k \,\partial\theta_{k'}$,
for $k,k'=1,\ldots,q$. Two additional regularity conditions are assumed
for Theorems \ref{thmTapexist} and \ref{thmTapose}:

\begin{longlist}[(A.13)]
\item[(A.12)]
$0<\inf_{n}\{\omega_n N_n^{-1/2}\} \leq\sup_{n}\{\omega_n N_n^{-1/2}\}
<\infty$, where $\omega_n=\omega$ is the threshold distance in the
tapering function (\ref{eqnKT}).

\item[(A.13)] There exists\vspace*{1pt} a nonincreasing function $\gamma_0$ with
$\int_{0}^{\infty}u^2\gamma_0(u)\,du<\infty$ such that $\max\{|\gamma
(u)|, |\gamma_k(u)|, |\gamma_{k,k'}(u)|\}\leq\gamma_0(u) $ for all $u
\in(0, \infty)$ and $1\leq k$, $k' \leq q$.
\end{longlist}

From (A.12), the threshold distance $\omega_n$ is bounded away from 0
and grows at the rate of $N_n^{1/2}$. The condition in (A.13) has to do
with the covariance function.
It can be shown that they hold for some of the commonly-used covariance
functions such as the Mat{\' e}rn class.
Details are given in Appendix~D of Chu et al. \cite{ChuZhuWang2011}.

%
%
\subsection{Consistency and asymptotic normality of PMLE$_{\mathrm T}$}

\begin{proposition}
\label{protapMLE} Under \textup{(A.1)--(A.7)} and
\textup{(A.12)--(A.13)}, the MLE$_{\mathrm T}$ $\Prefix
^{n}{\widehat{\mb{\eta}}}_{\mathrm{MLE}_{\mathrm T}}$ is asymptotically
normal with
\[
N_n^{1/2}(\Prefix^{n}{\widehat{\mb{\eta}}}_{\mathrm{MLE}_{\mathrm T}}-
\mb{\eta
}_{0})\stackrel{D}{\longrightarrow} N(\mb{0},\mathbf{J}(\mb{\eta}_{0})^{-1}).
\]
\end{proposition}

Proposition \ref{protapMLE} establishes the asymptotic normality of
MLE$_{\mathrm T}$. In particular, MLE and MLE$_{\mathrm T}$ have the same
limiting distribution. This implies that, under the regularity
conditions, covariance-tapered MLE achieves the same efficiency as MLE.
Thus, in Algorithm \ref{Algorithm1} for computing the OSE$_{\mathrm T}$, we may set the
initial parameter values to $\Prefix^{n}{\widehat{\mb{\eta}}}_{\mathrm{MLE}_{\mathrm{T}}}$.

%
\begin{theorem}
\label{thmTapexist} Under \textup{(A.1)--(A.9)} and
\textup{(A.12)--(A.13)}, there exists, with probability tending to one,
a local maximizer $\Prefix^{n}{\widehat{\mb{\eta}}}_{\mathrm T}$ of
$Q_{\mathrm T}(\mb{\eta})$ defined in (\ref{EqnTapQ}) such that
$\Vert\Prefix^{n}{\widehat{\mb{\eta}}}_{\mathrm T}-\mb{\eta}_{0}\Vert
=O_{p}(N_n^{-1/2}+a_{n})$. If,\vspace*{3pt} in addition, \textup{(A.10)--(A.11)}
hold, then $\Prefix^{n}{\widehat{\mb{\eta}}}_{\mathrm
T}=(\Prefix^{n}{\widehat{\mb {\beta}}}{}^{T}_{1,{\mathrm
T}},\Prefix^{n}{\widehat{\mb{\beta}}}{}^{T}_{2,{ \mathrm
T}},\Prefix^{n}{\widehat{\mb{\theta}}}{}^{T}_{\mathrm T})^{T}$
satisfies:
\begin{longlist}
\item Sparsity: $\Prefix^{n}{\widehat{\mb{\beta}}}_{2,{\mathrm
T}}=\mb
{0}$ with probability tending to 1.
\item Asymptotic normality:
\begin{eqnarray*}
&&N_n^{1/2}\{\mathbf{J}(\mb{\beta}_{10})+\mb{\Phi}_n(\mb{\beta}_{10})\}
[\Prefix^{n}{\widehat{\mb{\beta}}}_{1,{\mathrm T}}-\mb{\beta}_{10}+\{
\mathbf{J}(\mb{\beta}_{10})+\mb{\Phi}_n(\mb{\beta}_{10})\}^{-1}\mb{\phi
}_{n}(\mb{\beta}_{10})]
\\
&&\qquad\stackrel{D}{\longrightarrow} N(\mb{0},\mathbf{J}(\mb{\beta}_{10})), \\
&&N_n^{1/2}(\Prefix^{n}{\widehat{\mb{\theta}}}_{\mathrm T}-\mb{\theta}_{0})
\stackrel{D}{\longrightarrow} N(\mb{0},\mathbf{J}(\mb{\theta}_0)^{-1}),
\end{eqnarray*}
where $\mathbf{J}(\mb{\beta}_{10})$ and $\mb{\Phi}_n(\mb{\beta}_{10})$
consist of the first $s\times s$ upper-left submatrix of $\mathbf{J}(\mb
{\beta}_0)$ and $\mb{\Phi}_n(\mb{\beta}_0)$, respectively.
\end{longlist}
\end{theorem}

In Theorem \ref{thmTapexist}, PMLE$_{\mathrm T}$ is shown to be consistent,
sparse and asymptotically normal. In particular, PMLE$_{\mathrm T}$ has
the same asymptotic distribution as PMLE in Theorem \ref{thmexist}.
That is, PMLE$_{\mathrm T}$ achieves the same efficiency and oracle
property as PMLE asymptotically, yet in the mean time is more
computationally efficient.

%
%
\subsection{Consistency and asymptotic normality of OSE$_{\mathrm T}$}

%
\begin{theorem}
\label{thmTapose} Suppose that the initial value
$\Prefix^{n}{\mb{\eta}_{\mathrm T}^{(0)}}$ in Algorithm
\ref{Algorithm1} satisfies $\Prefix^{n}{\mb{\eta}_{\mathrm
T}^{(0)}}-\mb{\eta }_0=O_p(N_n^{-1/2})$. For the\vspace*{2pt} SCAD
penalty function, under \textup{(A.1)--(A.7)}, \textup{(A.10)} and
\textup{(A.12)--(A.13)}, the OSE$_{\mathrm T}$ $\Prefix
^{n}{\widehat{\mb{\eta}}}_{\mathrm{OSE}_{\mathrm
T}}=(\Prefix^{n}{\widehat{\mb{\beta }}}{}^{
T}_{1,{\mathrm{OSE}_{\mathrm T}}},
\Prefix^{n}{\widehat{\mb{\beta}}}{}^{T}_{2,{\mathrm{OSE}_{\mathrm
T}}},\break\Prefix ^{n}{\widehat{\mb{\theta}}}{}^{T}_{\mathrm{OSE}_{\mathrm
T}})^{T}$ satisfies:
\begin{longlist}
\item Sparsity: $\Prefix^{n}{\widehat{\mb{\beta}}}_{2,{
\mathrm{OSE}_{\mathrm T}}}=\mb{0}$ with probability tending to 1.
\item Asymptotic normality:
\begin{eqnarray*}
N_n^{1/2}(\Prefix^{n}{\widehat{\mb{\beta}}}_{1,{\mathrm{OSE}_{\mathrm T}}}-\mb{\beta
}_{10})&\stackrel{D}{\longrightarrow}& N(\mb{0},\mathbf{J}(\mb{\beta
}_{10})^{-1}),\\
N_n^{1/2}(\Prefix^{n}{\widehat{\mb{\theta}}}_{\mathrm{OSE}_{\mathrm T}}-\mb{\theta
}_{0})&\stackrel{D}{\longrightarrow}& N(\mb{0},\mathbf{J}(\mb{\theta}_0)^{-1}),
\end{eqnarray*}
where $\mathbf{J}(\mb{\beta}_{10})$ consists of the first $s\times s$
upper-left submatrix of $\mathbf{J}(\mb{\beta}_0)$.
\end{longlist}
\end{theorem}

Theorem \ref{thmTapose} establishes the asymptotic properties of
OSE$_{\mathrm T}$ under the SCAD penalty. In particular, OSE$_{\mathrm T}$
achieves the same limiting distribution as OSE of $\mb{\beta}_1$ and
$\mb{\theta}$ in Theorem \ref{thmose} and thus the same efficiency.
Furthermore, similar to Theorem \ref{thmose}, Theorem \ref{thmTapose}
holds for the class of penalty functions such that $p'_{\lambda_n}(\cdot
)=\lambda_np(\cdot)$ where $p'(\cdot)$ is continuous on $(0,\infty)$,
and there is some $\alpha>0$ such that $p'(\beta)=O(\beta^{-\alpha})$
as $\beta\rightarrow{0+}$, provided that $N_n^{(1+\alpha)/2}\lambda
_n\rightarrow\infty$ and $N_n^{1/2}\lambda_n\rightarrow0$.

\section{Numerical examples}\label{secnumeric}

%
%

\subsection{Simulation study}\label{subsecsimu}

We now conduct a simulation study to investigate the finite-sample
properties of OSE and OSE$_{\mathrm T}$. The spatial domain of interest is
assumed to be a square $[0,l]^2$ of side\vadjust{\goodbreak} lengths $l=5,10,15$. The
sample sizes are set to be $N=100, 400, 900$ for $l=5,10,15$,
respectively, with a fixed sampling density of 4. For regression, we
generate seven covariates that follow standard normal distributions
with a cross-covariate correlation of 0.5. The regression coefficients
are set to be $\mb{\beta}=(4,3,2,1,0,0,0)^T$. We standardize the
covariates to have mean 0 and variance 1 and standardize $\mathbf{y}$ to
have mean 0. Thus, there will be no intercept in the vector of
regression coefficients $\mb{\beta}$.
For spatial dependence, the error terms follow an
exponential covariance function $\gamma(d) = \sigma^2(1 - c) \exp( -
d/r)$, where $\sigma^2=9$ is the variance, $c=0.2$ is the nugget
effect and $r=1$ is the range parameter. For each choice of sample
size $N$, a total of $100$ data sets are simulated.

For each simulated data set, we compute OSE and OSE$_{\mathrm T}$ using
Algorithm~\ref{Algorithm1}. For OSE$_{\mathrm T}$, we consider different threshold values
for covariance tapering $\omega=l/2^k$ for $k=1,2,\ldots.$ We present
only the case of $\omega=l/4$ to save space. Our methods are compared
against several alternatives. Of particular interest is OSE under a
standard linear regression where spatial autocorrelation is unaccounted
for in the penalized loglikelihood function. This would be akin to PLS
under SCAD in
Wang and Zhu \cite{WangZhu2009} and will be referred to as OSE$_{
\mathrm{Alt}1}$. In addition, we modify the initialization step of Algorithm \ref{Algorithm1}
by using MLE under the true model which is unknown but assumed to be
known. This is an attempt to evaluate the effect of starting values and
will be referred to as OSE$_{\mathrm{Alt}2}$. Last, we consider a benchmark
case, referred to as OSE$_{\mathrm{Alt}3}$, where the true model is assumed
to be known and the MLE of the nonzero regression coefficients and the
covariance function parameters are computed. Our OSE and OSE$_{\mathrm T}$
will be compared against this benchmark to evaluate the oracle properties.

For each choice of sample size $N$, we first compute the average
numbers of correctly (C0) and incorrectly (I0) identified zero-valued
regression coefficients from OSE $\widehat{\mb{\beta}}_{\mathrm{OSE}}$ and
OSE$_{\mathrm T}$
$\widehat{\mb{\beta}}_{\mathrm{OSE}_{\mathrm T}}$, as well as those from OSE$_{
\mathrm{Alt}1}$ and OSE$_{\mathrm{Alt}2}$.
The true number of zero-valued regression coefficients is 3 as is
assumed in OSE$_{\mathrm{Alt}3}$.
Then, we compute means of the nonzero-valued OSE~$\widehat{\mb{\beta
}}_{1,\mathrm{OSE}}$ and
OSE$_{\mathrm T}$ $\widehat{\mb{\beta}}_{1,{\mathrm{OSE}_{\mathrm T}}}$, as well as the
corresponding covariance function
parameters $\widehat{\mb{\theta}}_{\mathrm{OSE}}$ and
$\widehat{\mb{\theta}}_{\mathrm{OSE}_{\mathrm T}}$. We estimate standard deviations
(SDs) of the parameter estimates using the information matrix.
The true SD is approximated by the median of the sample
SD (SDm) of the $100$ parameter estimates. The results are given in
Tables \ref{tabn100}--\ref{tabn900}.

%
\begin{table}[t!]
\caption{The average number of correctly identified 0 coefficients
(C0), average number of incorrectly identified 0 coefficients (I0),
mean, standard deviation (SD) and median estimated standard deviation
(SDm) under OSE, OSE$_{\mathrm T}$, OSE$_{\mathrm{Alt}1}$, OSE$_{\mathrm{Alt}2}$ and
OSE$_{\mathrm{Alt}3}$ for sample size $N=100$}
\label{tabn100}
\begin{tabular*}{\tablewidth}{@{\extracolsep{\fill}}lccd{2.2}ccc@{}}
\hline
\textbf{Method} & \textbf{Truth} & \textbf{OSE}
& \multicolumn{1}{c}{\textbf{OSE}$\bolds{_{\mathrm T}}$} &
\textbf{OSE}$\bolds{_{\mathrm{Alt}1}}$ & \textbf{OSE}$\bolds{_{
\mathrm{Alt}2}}$ & \textbf{OSE}$\bolds{_{\mathrm{Alt}3}}$ \\
\hline
C0 & 3\hphantom{.00} & 2.79 & 2.84 & 2.84 & 2.95 & 3.00 \\
I0 & & 0.06 & 0.10 & 0.32 & 0.06 & 0.00 \\
[4pt]
$\beta_1$ & 4.00 & 4.01 & 4.03 & 4.17 & 4.01 & 4.01 \\
SD && 0.28 & 0.29 & 0.39 & 0.27 & 0.27 \\
SDm && 0.26 & 0.27 & 0.36 & 0.26 & 0.26 \\
[4pt]
$\beta_2$ & 3.00 & 3.04 & 3.03 & 3.08 & 3.04 & 3.03\\
SD && 0.30 & 0.30 & 0.41 & 0.30 & 0.29 \\
SDm && 0.25 & 0.26 & 0.36 & 0.25 & 0.25 \\
[4pt]
$\beta_3$ & 2.00 & 1.94 & 1.97 & 2.00 & 1.94 & 1.93 \\
SD && 0.29 & 0.31 & 0.50 & 0.28 & 0.28 \\
SDm && 0.25 & 0.26 & 0.36 & 0.26 & 0.26 \\
[4pt]
$\beta_4$ & 1.00 & 1.02 & 1.03 & 0.78 & 1.03 & 1.02\\
SD && 0.35 & 0.40 & 0.55 & 0.33 & 0.26 \\
SDm && 0.24 & 0.24 & 0.26 & 0.24 & 0.26 \\
[4pt]
$r$ & 1.00 & 0.79 & 6.31 &-- &0.83 & 0.84 \\
SD && 0.54 & 2.14 &-- &0.57 & 0.57 \\
SDm && 0.48 & 17.65 &-- &0.51 & 0.51 \\
[4pt]
$c$ & 0.20 & 0.16 & 0.23 &-- &0.17 & 0.17\\
SD && 0.12 & 0.13 &-- &0.12 & 0.12 \\
SDm && 0.11 & 0.19 &-- &0.11 & 0.11 \\
[4pt]
$\sigma^2$ & 9.00 & 7.96 & 7.14 & 7.74 & 8.03 & 8.03 \\
SD && 2.28 & 1.53 & 2.06 & 2.36 & 2.36\\
SDm && 2.21 & 4.79 & 1.16 & 2.28 & 2.28\\
\hline
\end{tabular*}
\end{table}

In terms of variable selection, C0 tends to the true value 3 and I0
tends to 0, as the sample size $N$ increases, for OSE, OSE$_{\mathrm T}$,
OSE$_{\mathrm{Alt}1}$ and OSE$_{\mathrm{Alt}2}$. When the sample size is
relatively small ($N=100$), OSE$_{\mathrm{Alt}2}$ has the best performance
with the largest C0 and smallest I0, reflecting the effect of starting
values in Algorithm~\ref{Algorithm1}. But it is not practical, as we do not know what
the true model is in actual data analysis. OSE$_{\mathrm{Alt}1}$ assuming no
spatial dependence in the regression model seems to over-shrink the
regression coefficients. While C0 $=2.84$ is close to 3 under OSE$_{
\mathrm{Alt}1}$, I0 $=0.32$ is also large, compared to our OSE and OSE$_{\mathrm T}$. Between OSE and OSE$_{\mathrm T}$, it appears that C0 is slightly
better, but I0 is slightly worse for OSE$_{\mathrm T}$ than OSE.

%
\begin{table}[t!]
\caption{The average number of correctly identified 0 coefficients
(C0), average number of incorrectly identified 0 coefficients (I0),
mean, standard deviation (SD) and median estimated standard deviation
(SDm) under OSE, OSE$_{\mathrm T}$, OSE$_{\mathrm{Alt}1}$, OSE$_{\mathrm{Alt}2}$ and
OSE$_{\mathrm{Alt}3}$ for sample size $N=400$}
\label{tabn400}
\begin{tabular*}{\tablewidth}{@{\extracolsep{\fill}}lcccccc@{}}
\hline
\textbf{Method} & \textbf{Truth} & \textbf{OSE}
& \textbf{OSE}$\bolds{_{\mathrm T}}$ &
\textbf{OSE}$\bolds{_{\mathrm{Alt}1}}$ & \textbf{OSE}$\bolds{_{
\mathrm{Alt}2}}$ & \textbf{OSE}$\bolds{_{\mathrm{Alt}3}}$ \\
\hline
C0      & 3\hphantom{.00}  & 2.97 &   2.97 &2.97 &    2.98 &          3.00\\
I0      &  & 0.00 &     0.00 &0.01 &    0.00 &          0.00\\[4pt]
$\beta_1$ & 4.00 & 3.98 & 3.98 &3.98 & 3.99 & 3.99\\
SD && 0.14 & 0.14 &0.20 & 0.14 & 0.14\\
SDm && 0.13 & 0.13 &0.19 & 0.13 & 0.13\\
[4pt]
$\beta_2$ & 3.00 & 3.02 & 3.03 &3.03 & 3.02 & 3.02\\
SD && 0.14 & 0.14 &0.21 & 0.13 & 0.13\\
SDm && 0.13 & 0.13 &0.19 & 0.13 & 0.13\\
[4pt]
$\beta_3$ & 2.00 & 2.01 & 2.01 &2.01 & 2.01 & 2.01\\
SD && 0.12 & 0.12 &0.17 & 0.12 & 0.12\\
SDm && 0.13 & 0.13 &0.19 & 0.13 & 0.13\\
[4pt]
$\beta_4$ & 1.00 & 0.99 & 1.00 &0.96 & 1.00 & 1.00\\
SD && 0.12 & 0.12 &0.26 & 0.12 & 0.12\\
SDm && 0.13 & 0.13 &0.19 & 0.13 & 0.13\\
[4pt]
$r$ & 1.00 & 0.90 & 2.87 &-- & 0.90 & 0.90\\
SD && 0.29 & 4.08 & -- & 0.29 & 0.29\\
SDm && 0.25 & 5.24 &-- & 0.25 & 0.25\\
[4pt]
$c$ & 0.20 & 0.19 & 0.29 &-- & 0.19 & 0.19\\
SD && 0.06 & 0.07 &-- & 0.06 & 0.06\\
SDm && 0.05 & 0.11 &-- & 0.05 & 0.05\\
[4pt]
$\sigma^2$ & 9.00 & 8.70 & 8.25 &8.71 & 8.70 & 8.70\\
SD && 1.39 & 1.00 &1.37 & 1.39 & 1.39\\
SDm && 1.29 & 2.95 &0.63 & 1.29 & 1.29\\
\hline
\end{tabular*}
\end{table}

In terms of estimation of the nonzero regression coefficients, both
accuracy and precision improve as the sample size $N$ increases, for
all five OSE cases considered here. While the accuracy is similar
between OSE$_{\mathrm{Alt}1}$ and our OSE and OSE$_{\mathrm T}$, a striking
feature is the larger SD of OSE$_{\mathrm{Alt}1}$ when compared with our OSE
and OSE$_{\mathrm T}$, for all three sample sizes $N=100, 400, 900$. This
suggests that, by including spatial dependence directly in the
penalized likelihood function, we gain statistical efficiency in
parameter estimation. For the small sample size ($N=100$), SD based on
the information matrix without accounting for spatial dependence
appears to underestimate the true variation estimated by SDm.
Furthermore, the SD's of OSE and OSE$_{\mathrm T}$ tend to those in the
benchmark case OSE$_{\mathrm{Alt}3}$ as the sample size increases,
confirming the oracle properties established in Sections \ref
{secPMLEasymp} and \ref{sectaperasymp}. For 100 simulations, it takes
about 1 second, 30 seconds and 4 minutes per simulation for sample
sizes $N=$ 100, 400, 900, respectively.

Based on these simulation results, it may be tempting to consider using
OSE$_{\mathrm{Alt}1}$ to select variables and then OSE$_{\mathrm{Alt}3}$ for
parameter estimation when the sample size is reasonably large, as a
means of saving computational time. We contend that this is not
necessary, as our OSE or OSE$_{\mathrm T}$ enables variable selection and
parameter estimation simultaneously, at the similar computational cost.
Moreover, in practice, it is not always clear how large a sample size
at hand really is, as an effective sample size is influenced by factors
such as the strength of spatial dependence in the error process.

\subsection{Data examples}\label{subsecdata}

The first data example consists of January precipitation (inches per
24-hour period) on the log scale from 259 weather stations in the state
of Colorado \cite{ReichDavis2008}.
Candidate covariates are elevation, slope,
aspect and seven spectral bands from a MODIS satellite imagery (B1M
through B7M). It is of interest to investigate the relationship
between precipitation and these covariates.

%
\begin{table}[t!]
\caption{The average number of correctly identified 0 coefficients
(C0), average number of incorrectly identified 0 coefficients (I0),
mean, standard deviation (SD) and median estimated standard deviation
(SDm) under OSE, OSE$_{\mathrm T}$, OSE$_{\mathrm{Alt}1}$, OSE$_{\mathrm{Alt}2}$ and
OSE$_{\mathrm{Alt}3}$ for sample size $N=900$}
\label{tabn900}
\begin{tabular*}{\tablewidth}{@{\extracolsep{\fill}}lcccccc@{}}
\hline
\textbf{Method} & \textbf{Truth} & \textbf{OSE}
& \textbf{OSE}$\bolds{_{\mathrm T}}$ &
\textbf{OSE}$\bolds{_{\mathrm{Alt}1}}$ & \textbf{OSE}$\bolds{_{
\mathrm{Alt}2}}$ & \textbf{OSE}$\bolds{_{\mathrm{Alt}3}}$ \\
\hline
C0 & 3\hphantom{.00} & 3.00 & 3.00 & 3.00 & 3.00 & 3.00\\
I0 & & 0.00 & 0.00 & 0.00 & 0.00 & 0.00\\
[4pt]
$\beta_1$ & 4.00 & 4.00 & 4.01 &4.03 & 4.00 & 4.00\\
SD && 0.10 & 0.10 &0.13 & 0.10 & 0.10\\
SDm && 0.09 & 0.09 &0.13 & 0.09 & 0.09\\
[4pt]
$\beta_2$ & 3.00 & 3.01 & 3.01 &2.99 & 3.01 & 3.01\\
SD && 0.08 & 0.08 &0.12 & 0.08 & 0.08\\
SDm && 0.09 & 0.09 &0.13 & 0.09 & 0.09\\
[4pt]
$\beta_3$ & 2.00 & 1.98 & 1.99 &1.98 & 1.98 & 1.98\\
SD && 0.08 & 0.08 &0.11 & 0.08 & 0.08\\
SDm && 0.09 & 0.09 &0.13 & 0.09 & 0.09\\
[4pt]
$\beta_4$ & 1.00 & 1.00 & 1.00 &1.01 & 1.00 & 1.00\\
SD && 0.09 & 0.09 &0.13 & 0.09 & 0.09\\
SDm && 0.09 & 0.09 &0.13 & 0.09 & 0.09\\
[4pt]
$r$ & 1.00 & 0.94 & 1.44 &-- & 0.94 & 0.94\\
SD && 0.17 & 0.50 & -- & 0.17 & 0.17\\
SDm && 0.17 & 0.40 &-- & 0.17 & 0.17\\
[4pt]
$c$ & 0.20 & 0.19 & 0.25 &-- & 0.19 & 0.19\\
SD && 0.04 & 0.04 &-- & 0.04 & 0.04\\
SDm && 0.04 & 0.04 &-- & 0.04 & 0.04\\
[4pt]
$\sigma^2$ & 9.00 & 8.80 & 8.50 &8.80 & 8.80 & 8.80\\
SD && 0.90 & 0.74 &0.87 & 0.90 & 0.90\\
SDm && 0.85 & 1.15 &0.42 & 0.85 & 0.85\\
\hline
\end{tabular*}
\end{table}

We first fit a spatial linear model with an exponential covariance
function via maximum likelihood. The parameter estimates and their
standard errors in Table \ref{tabdata} suggest that the regression
coefficients for elevation, B1M, B4M, B6M and B7M are possibly
significant. Among the covariance function parameters, of most interest
is the range parameter, which is significantly different from zero.
This indicates that there is spatial autocorrelation among the errors
in the linear regression. Our OSE method selects elevation and B4M, and
shrinks all the other regression coefficients to zero. The covariance
function parameter estimates are close to the MLE. For comparison, we
fit a~standard linear regression with i.i.d. errors and the corresponding
OSE$_{\mathrm{Alt}1}$ selects slope and aspect in addition to elevation and
B4M. However, the regression coefficients for slope and aspect do not
appear to be significant.

%
\begin{table}
\caption{Regression coefficient estimates and standard deviations (SD)
using maximum likelihood (MLE) and one-step sparse estimation (OSE)
under a spatial linear model with an exponential covariance function
for the Gaussian error process, as well as OSE under a~standard linear
model with i.i.d. errors (OSE$_{\mathrm{Alt}1}$)}
\label{tabdata}
\begin{tabular*}{\tablewidth}{@{\extracolsep{\fill}}ld{2.3}cd{2.3}cd{2.3}c@{}}
\hline
\textbf{Terms} & \multicolumn{1}{c}{\textbf{MLE}}
& \multicolumn{1}{c}{\textbf{SD}} & \multicolumn{1}{c}{\textbf{OSE}}
& \multicolumn{1}{c}{\textbf{SD}} & \multicolumn{1}{c}{\textbf{OSE}$\bolds{_{\mathrm{Alt}1}}$}
& \multicolumn{1}{c@{}}{\textbf{SD}} \\
\hline
\multicolumn{7}{@{}c@{}}{Regression coefficients} \\[4pt]
Elevation & 0.305 & 0.055 & 0.228 & 0.054 & 0.195 & 0.044 \\
Slope & 0.016 & 0.026 & \multicolumn{1}{c}{--} & \multicolumn{1}{c}{--} & 0.035 & 0.040 \\
Aspect &-0.004 & 0.022 & \multicolumn{1}{c}{--} & \multicolumn{1}{c}{--} & 0.032 & 0.034 \\
B1M & 0.214 & 0.157 & \multicolumn{1}{c}{--} & \multicolumn{1}{c}{--} & \multicolumn{1}{c}{--} & \multicolumn{1}{c@{}}{--} \\
B2M & 0.058 & 0.064 & \multicolumn{1}{c}{--} & \multicolumn{1}{c}{--} & \multicolumn{1}{c}{--} & \multicolumn{1}{c@{}}{--} \\
B3M & 0.017 & 0.109 & \multicolumn{1}{c}{--} & \multicolumn{1}{c}{--} & \multicolumn{1}{c}{--} & \multicolumn{1}{c@{}}{--} \\
B4M & -0.404& 0.183 & -0.089 & 0.034 & -0.264 & 0.045 \\
B5M & 0.043 & 0.089 & \multicolumn{1}{c}{--} & \multicolumn{1}{c}{--} & \multicolumn{1}{c}{--} & \multicolumn{1}{c@{}}{--} \\
B6M & -0.162& 0.116 & \multicolumn{1}{c}{--} & \multicolumn{1}{c}{--} & \multicolumn{1}{c}{--} & \multicolumn{1}{c@{}}{--} \\
B7M & 0.172 & 0.098 & \multicolumn{1}{c}{--} & \multicolumn{1}{c}{--} & \multicolumn{1}{c}{--} & \multicolumn{1}{c@{}}{--} \\
[4pt]
\multicolumn{7}{@{}c@{}}{Covariance function parameters}\\[4pt]
Range & 0.967 & 0.368 & 1.043 & 0.417 & \multicolumn{1}{c}{--} & \multicolumn{1}{c@{}}{--} \\
Nugget & 0.183 & 0.061 & 0.196 & 0.064 & \multicolumn{1}{c}{--} & \multicolumn{1}{c@{}}{--} \\
$\sigma^2$& 0.287 & 0.067 & 0.304 & 0.074 & 0.289 & 0.026 \\
\hline
\end{tabular*}\vspace*{6pt}
\end{table}

In addition, we apply our method to the whiptail lizard data as
described in Section \ref{secintro}. There are 148 sites, and the
response variable is the abundance of lizards at each site. There are
26 covariates regarding location, vegetation, flora, soil and ants.
Hoeting et al. \cite{HoetingDavisMertonThompson2006} considered only 6
covariates
after a separate prescreening procedure, and selected 2 covariates in
their final model. In this paper, we consider all 26 covariates
simultaneously, and interestingly reach the same final model. For
details of the results, see Section 6.2 in Chu et al.
\cite{ChuZhuWang2011}.

\begin{appendix}\label{app}
\section*{Appendix: Technical details}
For ease of notation, we suppress $n$ in $\Prefix^{n}{t_{kk'}}$,
$\Prefix^{n}{a_{kk'}}$, $\Prefix^{n}{\mb{\Gamma}}$, $\mathbf{I}_n$,
$\mathbf{A}_n$, $\Prefix^{n}{\widehat{\mb{\eta}}}$,
$\Prefix^{n}{\widehat{\mb{\beta}}}$ and
$\Prefix^{n}{\widehat{\mb{\theta}}}$. The detailed proofs of all lemmas
and theorems are given in Chu et al. \cite{ChuZhuWang2011}.

\subsection{Asymptotic properties of PMLE and OSE}\label{appA1}

\begin{lemma}\label{lemlemma1}
Under \textup{(A.1)--(A.7)}, for any given $\mb{\eta}\in\mathbb{R}^p\times\Omega
$, we have, as $n\rightarrow\infty$,
\[
N_n^{-1/2}\ell'(\mb{\eta})\stackrel{D}{\longrightarrow} N(\mb{0},\mathbf
{J}(\mb{\eta})),\qquad  N_n^{-1}\ell''(\mb{\eta})\stackrel
{P}{\longrightarrow} -\mathbf{J}(\mb{\eta}),
\]
where
$\mathbf{J}(\mb{\eta})=\operatorname{diag}\{\mathbf{J}(\mb{\beta}),\mathbf{J}(\mb
{\theta})\}$.
\end{lemma}
%
%
\begin{Remark*}
Lemma \ref{lemlemma1} establishes the asymptotic behavior of the
first-order and the second-order derivatives of the log-likelihood
function $\ell(\mb{\eta})$, scaled by $N_n^{-1/2}$ and $N_n^{-1}$,
respectively. In addition, by Theorem 2 of Mardia and Marshall \cite
{MardiaMarshall1984}, $\widehat{\mb{\eta}}_{\mathrm{MLE}}$ is consistent and
asymptotically normal with $\Vert\widehat{\mb{\eta}}_{\mathrm{MLE}}-\mb{\eta
}_0\Vert=O_p(N_n^{-1/2})$ and $N_n^{1/2}(\widehat{\mb{\eta}}_{\mathrm{
MLE}}-\mb{\eta}_0)\stackrel{D}{\longrightarrow} N(\mb{0},\mathbf{J}(\mb{\eta
}_0)^{-1})$. Moreover,\vspace*{2pt} for a random vector $\mb{\eta}^*$, such that
$\Vert\mathbf{I}(\mb{\eta})^{1/2}(\mb{\eta}^*-\mb{\eta}) \Vert= O_p(1)$,
by Theorem 2 of Mardia and Marshall \cite{MardiaMarshall1984}, we have
$ N_n^{-1}\ell''(\mb{\eta}^*)\stackrel{P}{\longrightarrow}
-\mathbf{J}(\mb
{\eta})$. These results will be used repeatedly in the proof of
Theorems \ref{thmexist} and \ref{thmose}.
\end{Remark*}
%
%
\begin{pf*}{Proof of Theorem \ref{thmexist}}
The proof follows from Lemma \ref{lemlemma1} and arguments extended
from Theorems 1 and 2 in Fan and Li \cite{FanLi2001}. See details in
Chu et al. \cite{ChuZhuWang2011}.
\end{pf*}
%
%
\begin{pf*}{Proof of Theorem \ref{thmose}}
The proof follows from Lemma \ref{lemlemma1} and arguments extended
from Theorem 5 in Zou and Li \cite{ZouLi2008}. See details in
Chu et al. \cite{ChuZhuWang2011}.
\end{pf*}

%
%
\subsection{Asymptotic properties of PMLE$_{\mathrm T}$ and
OSE$_{\mathrm T}$}\label{appA2}

Let $|A|$ denote the cardinality of a discrete set $A$. Let $\mu_{1,
\mathrm T} \leq\cdots\leq\mu_{N_n,\mathrm T}$ denote the eigenvalues of tapered
covariance matrix $\mb{\Gamma}_{\mathrm T}$. Let $\mu^{k}_{l,\mathrm T}$ denote
the eigenvalues of $\mb{\Gamma}_{k,\mathrm T}$ such that $|\mu^{k}_{1,\mathrm
T}| \leq\cdots\leq|\mu^{k}_{N_n,\mathrm T}|$ and let $\mu^{kk'}_{l,\mathrm
T}$ denote the eigenvalues of $\mb{\Gamma}_{kk',\mathrm T}$ such that $|\mu
^{kk'}_{1,\mathrm T}| \leq\cdots\leq|\mu^{kk'}_{N_n,\mathrm T}|$. For a
matrix $\mathbf{A}$, we let $\mu_{\min}(\mathbf{A})$ denote the minimum
eigenvalue of $\mathbf{A}$. Also, recall that $t_{kk',\mathrm
T}=\operatorname{tr}(\mb
{\Gamma}_{\mathrm T}^{-1}\mb{\Gamma}_{k,\mathrm T}\mb{\Gamma}_{\mathrm
T}^{-1}\mb
{\Gamma}_{k',\mathrm T})$.
%
%
\begin{lemma}\label{lemlemma2}
Under \textup{(A.12)--(A.13)}, we have:
\begin{eqnarray*}
\mbox{\textup{(i)}\hphantom{ii}\quad\hspace*{18.9pt}} \Vert\mb{\Gamma}-\mb{\Gamma}_{\mathrm
T}\Vert_\infty&=&O(N_n^{-1/2});\qquad
\mbox{\textup{(ii)}\quad} \Vert\mb{\Gamma}_k-\mb{\Gamma}_{k,\mathrm T}\Vert_\infty
=O(N_n^{-1/2});\\
\mbox{\textup{(iii)}\quad} \Vert\mb{\Gamma}_{kk'}-\mb{\Gamma}_{kk',\mathrm T}\Vert_\infty
&=&O(N_n^{-1/2}).
\end{eqnarray*}
\end{lemma}
%
%
%
%
%
%
%
\begin{Remark*}
Lemma \ref{lemlemma2} establishes that the order of the difference
between the covariance matrix $\mb{\Gamma}$ and the tapered covariance
matrix $\mb{\Gamma}_{\mathrm T}$ is $N_n^{-1/2}$, as well as that of the
first-order and the second-order derivatives of the covariance
matrices. These results are used when establishing Lemma~\ref{lemlemma3}.
\end{Remark*}
%
%
\begin{lemma}\label{lemlemma3}
Under \textup{(A.1)--(A.4)}, \textup{(A.6)} and \textup{(A.12), (A.13)}, we
have:
\begin{longlist}[(C.4)]
\item[(C.1)] $\hspace*{-0.5pt}\lim_{n \rightarrow\infty} \mu_{N_n,\mathrm T} \hspace*{-0.1pt}=\hspace*{-0.1pt}
C<\infty,\hspace*{-0.3pt}
{\lim_{n \rightarrow\infty}} |\mu_{N_n,\mathrm T}^{k}| \hspace*{-0.1pt}=\hspace*{-0.1pt} C_{k}<\infty,\hspace*{-0.3pt}
{\lim_{n \rightarrow\infty}} |\mu_{N_n,\mathrm T}^{kk'}| \hspace*{-0.3pt}= C_{kk'}<\infty$
for any $k,k'=1,\ldots,q$.

\item[(C.2)] For $k=1,\ldots,q$, $\Vert\mb{\Gamma}_{k,\mathrm T}\Vert
_F^{-2}=O(N_n^{-1/2-\delta})$, for some $\delta>0$.

\item[(C.3)] $\Vert\mb{\Gamma}_{\mathrm T}^{-1}\Vert_{s}<C_0<\infty$.

\item[(C.4)] For any $k,k'=1,\ldots,q$, $a_{kk',\mathrm T}=\lim\{t_{kk',\mathrm
T}(t_{kk,\mathrm T}t_{k'k',\mathrm T})^{-1/2}\}$ exists and is equal to
$a_{kk'}=\lim\{t_{kk'}(t_{kk}t_{k'k'})^{-1/2}\}$. That is, $\mathbf{A}_{\mathrm T}=(a_{kk',\mathrm T})_{k,k'=1}^q=\mathbf{A}=(a_{kk'})_{k,k'=1}^q$ and is nonsingular.
\end{longlist}
\end{lemma}

\begin{Remark*}
Conditions (C.1)--(C.4) are the covariance tapering counterparts of (A.2),
(A.3)(i), (A.4)(i) and (A.6). Together with (A.5), they yield
Proposition~\ref{protapMLE}. In fact, Lemmas \ref{lemlemma2} and \ref
{lemlemma3} hold for other tapering functions such as truncated
polynomial functions of $d/\omega$ with constant term equal to~$1$ when
$d < \omega$, and $0$ otherwise \cite{Wendland1995}. Furthermore,
(A.12) can be weakened to $0<\inf_{n}\{\omega_n N_n^{-1/2+\tau}\} \leq
\sup_{n}\{\omega_n N_n^{-1/2+\tau}\}<\infty$, with $\tau<\min\{
1/2,\delta\}$.
\end{Remark*}
\begin{lemma}\label{lemlemma4}
Under \textup{(A.1)--(A.7)} and \textup{(A.12)--(A.13)}, for any given
$\mb{\eta}\in \mathbb{R}^p\times\Omega$, we have
\[
N_n^{-1/2}\ell'_{\mathrm T}(\mb{\eta}) \stackrel{D}{\longrightarrow}
N(\mb
{0}, \mathbf{J}(\mb{\eta}))
\quad \mbox{and}\quad
N_n^{-1}\ell''_{\mathrm T}(\mb{\eta}) \stackrel{P}{\longrightarrow} -\mathbf
{J}(\mb{\eta}),
\]
where recall that $\mathbf{J}(\mb{\eta})=\operatorname{diag}\{\mathbf{J}(\mb{\beta}),\mathbf
{J}(\mb{\theta})\}$.
\end{lemma}
%
%
%
\begin{Remark*}
Lemma \ref{lemlemma4} establishes the asymptotic behavior of the
first-order and the second-order derivatives of the covariance-tapered
log-likeli\-hood function $\ell_{\mathrm T}(\mb{\eta})$. The rates of
convergence and the limiting distributions are the same as those for
the log-likelihood function. As in Lemma \ref{lemlemma1}, it follows
that MLE$_{\mathrm T}$ $\widehat{\mb{\eta}}_{\mathrm{MLE}_{\mathrm T}}$ is consistent and
asymptotically normal, as is given in Proposition \ref{protapMLE}.
These results will be used to establish Theorems~\ref{thmTapexist} and~%
\ref{thmTapose} and play the same role as Lemma~\ref{lemlemma1} when
showing Theorems \ref{thmexist} and \ref{thmose}.
\end{Remark*}
%
%
\begin{pf*}{Proof of Proposition \ref{protapMLE}}
From Lemma \ref{lemlemma3}, (C.1)--(C.4) are satisfied. Together with
(A.5), the regularity conditions of Theorem 2 of Mardia and Marshall
\cite{MardiaMarshall1984} hold. Thus, the result in Proposition \ref
{protapMLE} follows.
\end{pf*}
%
%
\begin{pf*}{Proof of Theorem \ref{thmTapexist}}
The proof of Theorem \ref{thmTapexist} is similar to that of
Theorem \ref{thmexist}. The main differences are that the parameter
estimates $ \widehat{\mb{\eta}}_{\mathrm{PMLE}}$, log-likelihood function
$\ell(\mb{\eta})$ and penalized log-likelihood $Q(\mb{\eta})$ are
replaced with their covariance-tapered counterparts $\widehat{\mb{\eta
}}_{\mathrm{PMLE}_{\mathrm T}}$, $\ell_{\mathrm T}(\mb{\eta})$ and $Q_{\mathrm T}(\mb{\eta})$,
respectively. Furthermore, we replace the results from Lemma \ref
{lemlemma1} with those from Lemma \ref{lemlemma4}, which holds due to
Lemmas \ref{lemlemma2} and \ref{lemlemma3} under the additional
assumptions (A.12) and (A.13).\vadjust{\goodbreak}
\end{pf*}
%
%
\begin{pf*}{Proof of Theorem \ref{thmTapose}}
The proof of Theorem \ref{thmTapose} is similar to that of Theorem \ref
{thmose}, but we replace the parameter estimates $ \widehat{\mb{\eta
}}_{\mathrm{OSE}}$, log-likelihood function $\ell(\mb{\eta})$ and
$Q^*(\mb
{\beta})$ with their covariance-tapered counterparts $\widehat{\mb{\eta
}}_{\mathrm{OSE}_{\mathrm T}}$, $\ell_{\mathrm T}(\mb{\eta})$ and $Q^*_{\mathrm T}(\mb{\beta
})$, respectively. As before, we replace the results from Lem\-ma~\ref
{lemlemma1} with those from Lemma \ref{lemlemma4}, where the additional
conditions (A.12) and~(A.13) are assumed and Lemmas \ref
{lemlemma2} and \ref{lemlemma3} are applied.
\end{pf*}
\end{appendix}

\section*{Acknowledgments}

We are grateful to the Editor, Associate Editor and three anonymous
referees for their helpful and constructive comments. We thank Drs.
Jennifer Hoeting and Jay Ver Hoef for providing the lizard data.


%

\printaddresses

\end{document}